\documentclass[fleqn,twoside,twocolumn,nofootinbib]{revtex4} 
\usepackage{ujp} 
\begin{document}
\title[MOLECULAR STRUCTURE AND INTERACTIONS]
{MOLECULAR STRUCTURE AND\\ INTERACTIONS  OF NUCLEIC
ACID COMPONENTS\\ IN\, NANOPARTICLES:\, \boldmath$AB$\, $INITIO$\, CALCULATIONS}%
\author{YU.V. RUBIN}
\affiliation{B. Verkin Institute for Low Temperature Physics and
Engineering,\\ Nat.
Acad. of Sci. of Ukraine}
\address{47, Lenin Ave., Kharkiv 61103, Ukraine}
\email{rubin@ilt.kharkov.ua}
\author{L.F. BELOUS}%
\affiliation{B. Verkin Institute for Low Temperature Physics and
Engineering,\\ Nat.
Acad. of Sci. of Ukraine}%
\address{47, Lenin Ave., Kharkiv 61103, Ukraine}%
\email{rubin@ilt.kharkov.ua}%
\udk{577.3}  \razd{\secix}

\setcounter{page}{723}%
\maketitle

\begin{abstract}
Self-associates of nucleic acid components (stacking trimers and
tetramers of the base pairs of nucleic acids) and short fragments of
nucleic acids are nanoparticles (linear sizes of these particles are
more than 10 {\AA}). Modern quantum-mechanical methods and softwares
allow one to perform {\it ab initio} calculations of the systems
consisting of 150--200 atoms with enough large basis sets (for
example, 6-31G*). The aim of this work is to reveal the
peculiarities of molecular and electronic structures, as well as the
energy features of nanoparticles of nucleic acid components.

We had carried out {\it ab initio} calculations of the molecular
structure and interactions in the stacking dimer, trimer, and
tetramer of nucleic base pairs and in the stacking (TрG)(AрC) dimer
and (TрGpC) (АрCpG) trimer of nucleotides, which are small DNA
fragments.

The performed calculations of molecular structures of dimers and
trimers of nucleotide pairs showed that the interplanar distance in
the structures studied is equal to 3.2~{\AA} on average, and the helical
angle in a trimer is approximately equal to $30^{\circ}.$ The
distance between phosphor atoms in neighboring chains is 13.1~{\AA}.
For
dimers and trimers under study, we calculated the horizontal interaction energies.

The analysis of interplanar distances and angles between nucleic
bases and their pairs in the calculated short oligomers of nucleic
acid base pairs (stacking dimer, trimer, and tetramer) has been
carried out. Studies of interactions in the calculated short
oligomers showed a considerable role of the cross interaction in the
stabilization of the structures. The contribution of cross
interactions to the horizontal interactions grows with the length of
an oligomer. Nanoparticle components get electric charges in
nanoparticles. Longwave low-intensity bands can appear in the
electron spectra of nanoparticles.
\end{abstract}

\section{Introduction}

Self-associates of nucleic acids (stacking trimers and tetramers of
nucleic acid base pairs) and short NA fragments are nanoparticles
((linear sizes of these particles are more than 10~{\AA}). Modern
quantum-mechanical methods and programs allow one to perform the
{\it ab initio} calculations of systems consisting of 150--200 atoms
with a rather large basis (6-31G*, for example) [1--3].

Earlier, the {\it ab initio} calculations were carried out for
nucleic acid base pairs and for dimers of nucleic acid base pairs
[4--10]. But the sizes of these structures are less than 10~nm, and
these structures cannot be considered as nanoparticles. Work [11]
presents the results of calculations for interaction energies of an
oligomer including 7 pairs of nucleic acid bases. But, in those
calculations, the oligomer geometry was not optimized, being taken
from NMR data.

Some time ago, we published the results of calculations for the
molecular structure of a stacking trimer of nucleic acid base pairs
[12].

The aim of the present work is to reveal the peculiarities of molecular and
electronic structures and energy characteristics of nanoparcticles of
nucleic acid components.

\section{Subjects and Methods of Calculations}

We performed {\it ab initio} calculations of the molecular and
electronic structures of short DNA fragments -- stacking (TрG)(AрC)
dimer and (TрGpC) (АрCpG) trimer of nucleotide pairs with four
Na$^{+}$ ions and without Na$^{+},$ as well as stacking dimer,
trimer, and tetramer of nucleic acid base pairs.

Geometries of the stacking (AC)(TG) dimer of base pairs, (CAC)(GTG)
trimer, (ACAC)(TGTG) tetramer, and DNA short fragments were
optimized by M06-2x functional of the DFT method [13]. As was shown
in [14], this functional describes well the stacking interaction of
nucleic acid components. Then, the geometries of dimers and trimers
of the base pairs were re-optimized at MP2 level of theory. 6-31/G*
basis set was used for the oligomer geometry optimization. The
analysis of bond lengths in a dimer calculated by the above methods
revealed a difference of 0.1{\%} in the values of bond lengths.

\begin{figure}
\includegraphics[width=7.5cm]{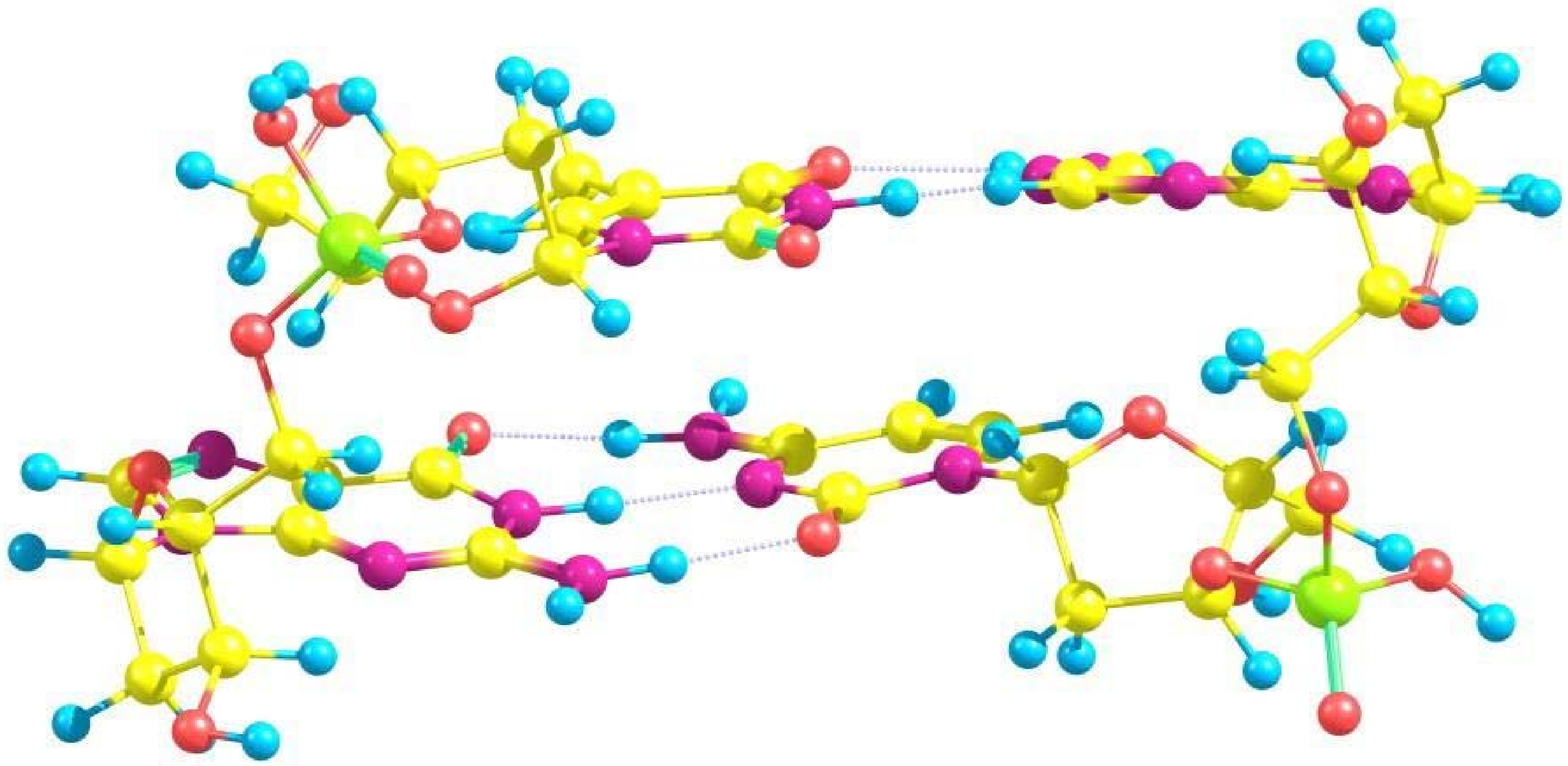}\\
TpA...GpC\\
\includegraphics[width=7.5cm]{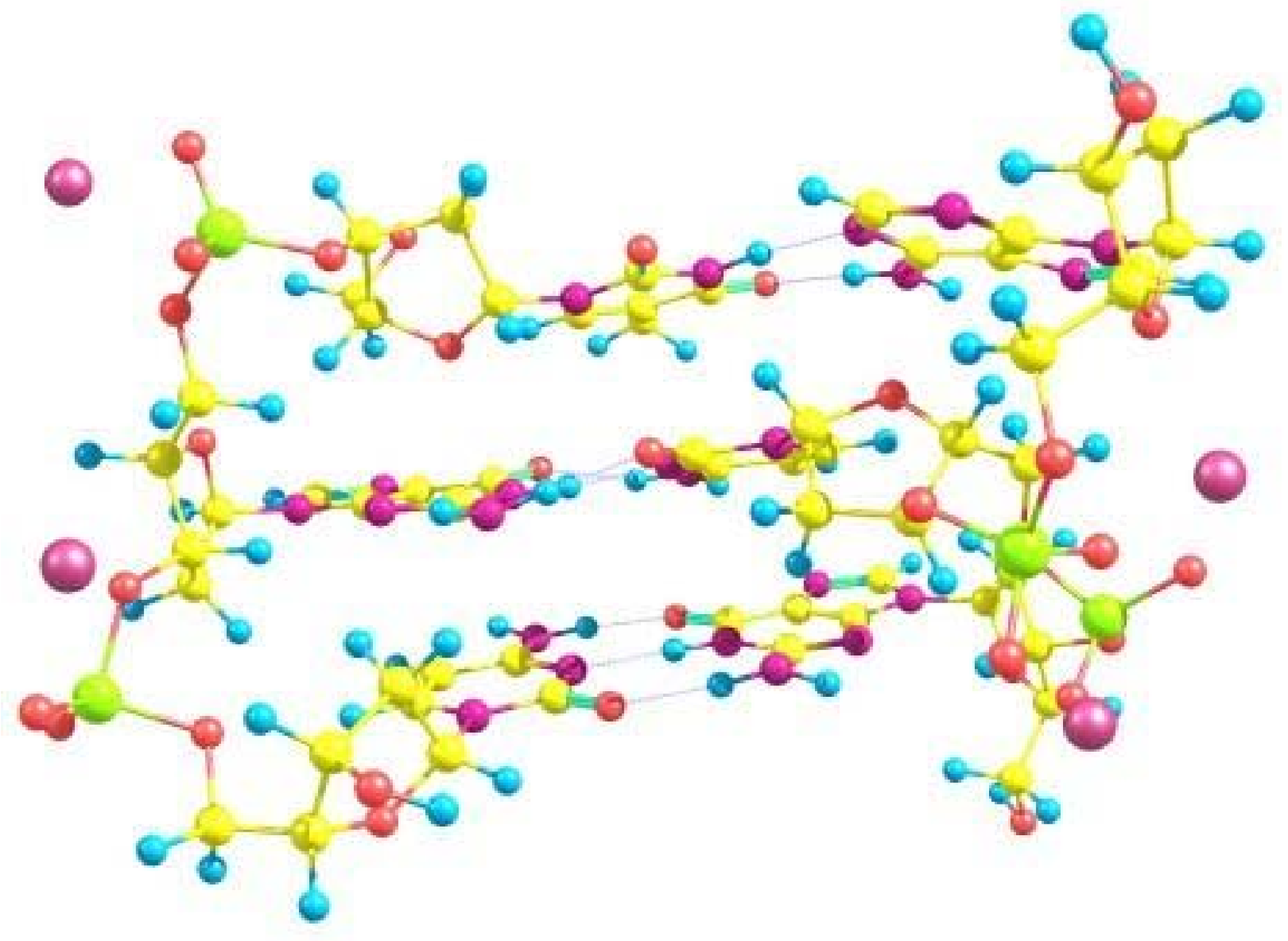}\\
Tp...Ap\\
Gp...Cp\\
\hspace{1.2cm}C...G + 4Na$^+$\\
 \vskip-3mm\caption{ Optimized structures of stacking dimer and
trimer of nucleotides }
\end{figure}

The analysis of the vibrational spectra of short oligomers showed the absence of
imaginary frequencies in the calculated spectra of dimers, trimers, and
tetramers. The interaction energies in these structures were determined with the
typical scheme described in [4]. According to this scheme, the interaction
energy in a complex is equal to the difference between the total energy
of the complex and the sum of total energies of monomers. These calculations
were performed at MP2 level of theory, BSSE being taken into account and
6-31+G* basis set being applied. The charges on atoms were calculated by
the Merz--Kolman method [15]. The calculation of excited state energies had been
carried out by the TD DFT method. Here, we used NWChem [1], Gamess [2], and
Gaussian 03 [3] programs. The visualization of the results of calculations is performed
by the Chem Craft Program [16].

\section{Results and Discussion}

\subsection{Molecular structure of DNA fragments}

When considering the calculation results for molecular structures of
a dimer and a trimer of nucleotide pairs (Fig. 1), it should be
noted that the calculated (АрCpG) trimer is a codon and is able to
code tryptophan aminoacid [17]. Calculations of the molecular
structure revealed (Table 1) that, in the structures studied, the
average value of interplanar spacing is 3.2 {\AA}, the helicity
angle in a trimer is $30^{\circ},$ and distances between phosphorus
atoms in neighbor chains are 13.1 {\AA}, respectively. Intervals
between extreme hydrogen atoms in nucleotide pairs and lengths of
hydrogen bonds are 15.8 and 1.9 {\AA} on the average, respectively.
Comparison of the calculated covalent bond lengths in a dimer and a
trimer with the X-ray structure analysis data revealed their
difference of about 2{\%}. We calculated the horizontal interactions
in the dimer and the trimer under study. These energies are 40 and
69 kcal/mol, respectively. These energies are larger than the total
energy of hydrogen bonds, and this is related to the contribution of
the cross interactions to the energy of horizontal interactions in
oligomers of nucleic acid components (as will be shown
below).\looseness=1

\begin{table}[b]
\noindent\caption{Geometry Parameters of Nucleic
Acid}\vskip3mm\tabcolsep9.0pt
\noindent{\footnotesize\begin{tabular}{l c c }
 \hline \multicolumn{1}{c}
{\rule{0pt}{9pt}Parameters} & \multicolumn{1}{|c}{Dimer}&
\multicolumn{1}{|c}{Trimer}\\%
\hline%
Interplanar distance, {\AA} & 3.2 & 3.2 \\%
Helical angle, degrees & 10 & 30  \\%
Distance between P atoms in & &   \\%
adjacent chains, {\AA} &  13.1 & 13.2 \\%
Distance between Na$^+$ ions in &  &  \\%
adjacent chains, {\AA} &  & 15.8  \\%
Horizontal interaction energies, &   &    \\%
kcal/mol&--40.2&--69.1*\\
\hline \multicolumn{3}{l}{*\,This calculation is performed for
a trimer without Na$^{+}$ ions}\\
\end{tabular}}
\end{table}

\subsection{Molecular structures of nucleic base pair oligomers}

Figure 2 presents the optimized geometries calculated for the
stacking (AC)(TG) dimer, (CAC)(GTG) trimer, and (ACAC)(TGTG)
tetramer of nucleic acid base pairs.

Calculations of the stacking dimer, trimer, and tetramer revealed
the parameters of structures presented in Table 2.

\begin{table}[b]
\noindent\caption{Structural parameters of calculated stacking
dimer, trimer, and tetramer}\vskip3mm\tabcolsep2.5pt
\noindent{\footnotesize\begin{tabular}{l r c r c r c}
 \hline \multicolumn{1}{c}
{\rule{0pt}{9pt}Structural} & \multicolumn{2}{|c}{(AC)(TG)}&
\multicolumn{2}{|c}{(CAC)(GTG)}&
\multicolumn{2}{|c}{(ACAC)(TGTG)}\\%
 \multicolumn{1}{c} {parameter} & \multicolumn{2}{|c}{}&
\multicolumn{2}{|c}{}&
\multicolumn{2}{|c}{}\\%
\hline%
Helical twist, & $0^{\circ}$ && $0^{\circ}$& CG1--CG3 & $5^{\circ}$& AT1--CG2 \\%
degree & & & $10^{\circ}$& CG1--AT2 &$2^{\circ}$& CG2--AT3 \\%
 &  && $10^{\circ}$& AT2--CG3 &$6^{\circ}$& AT3--CG4 \\[2mm]%
Propeller twist, & $15^{\circ}$& AT & $5^{\circ}$& CG1 &$2^{\circ}$& AT1 \\%
degree & $20^{\circ}$& GC &$0^{\circ}$& AT &$11^{\circ}$& CG2\\%
 &  & &$3^{\circ}$ &CG3 &$0^{\circ}$& AT3 \\%
 &  &&  && $8^{\circ}$& CG4\\[2mm]%
Dihedral angle & $175^{\circ}$& AT & $178^{\circ}$& CG1 & $161^{\circ}$& AT1 \\%
between bases, & $176^{\circ}$& GC & $180^{\circ}$& AT & $176^{\circ}$& CG2 \\%
degree &  && $177^{\circ}$& CG3 & $179^{\circ}$& AT3 \\%
 &  & && & $-170^{\circ}$& CG4 \\%
\hline
\end{tabular}}
\end{table}

The interplanar distance in the structures is 3.2~{\AA} on the
average.

The value of helical twist in the structures lies in
the $0-10^{\circ} $ interval (at first, $40^{\circ} $ angle was specified).

The value of propeller twist between bases of the studied structures
is in the 3--20$^{\circ} $ interval. The value of propeller twist in base pairs of the
dimer is larger than that in the trimer.

Base pairs in the structures are nonplanar. The value of
dihedral angle between base planes changes within
160--180$^{\circ}.$

Middle pairs in the trimer and in the tetramer are more planar (the dihedral angle and
the propeller one are equal to 0) than outside ones. On the whole, pairs in
the trimer are more planar than those in the dimer. This is an important peculiarity of
the molecular structure calculated for the trimer and the tetramer.

\subsection{Interactions in stacking oligomers of base pairs}

We performed calculations of the vertical and horizontal
interactions in the oligomers under study. The vertical interaction
energies between AT and CG pairs are equal to $-12.02$~kcal/mol in a
dimer and $-10.98$ kcal/mol (between high and middle pairs) or
$-11.0$ kcal/mol (between middle and down pairs) in a trimer,
respectively. These data correlate with results of work [2], in
which the parameters of the vertical interactions in stacking dimers
of base pairs were calculated by the complete basis set
extrapolation.

It is of interest to evaluate the energies of horizontal interactions in
the studied structures. Table 3 presents the calculation results for
the energies of horizontal interactions in the oligomers and
the sums of hydrogen bond energies in AT and GC pairs forming these
oligomers. Our calculations of the energies of hydrogen bonds in
isolated pairs of AT and GC gave $-13.83$ kcal/mol for AT and
$-26.98$ kcal/mol for GC pairs. As is seen from Table 3, the energies of
horizontal interactions exceed the arithmetic sum of the energies of
hydrogen bonds in isolated AT and GC pairs. These overridings are
equal to $-2.7$ kcal/mol, $-9.39$ kcal/mol, and $-13.1$ kcal/mol for
the dimer, trimer, and tetramer, respectively.

\begin{figure}
\includegraphics[width=5cm]{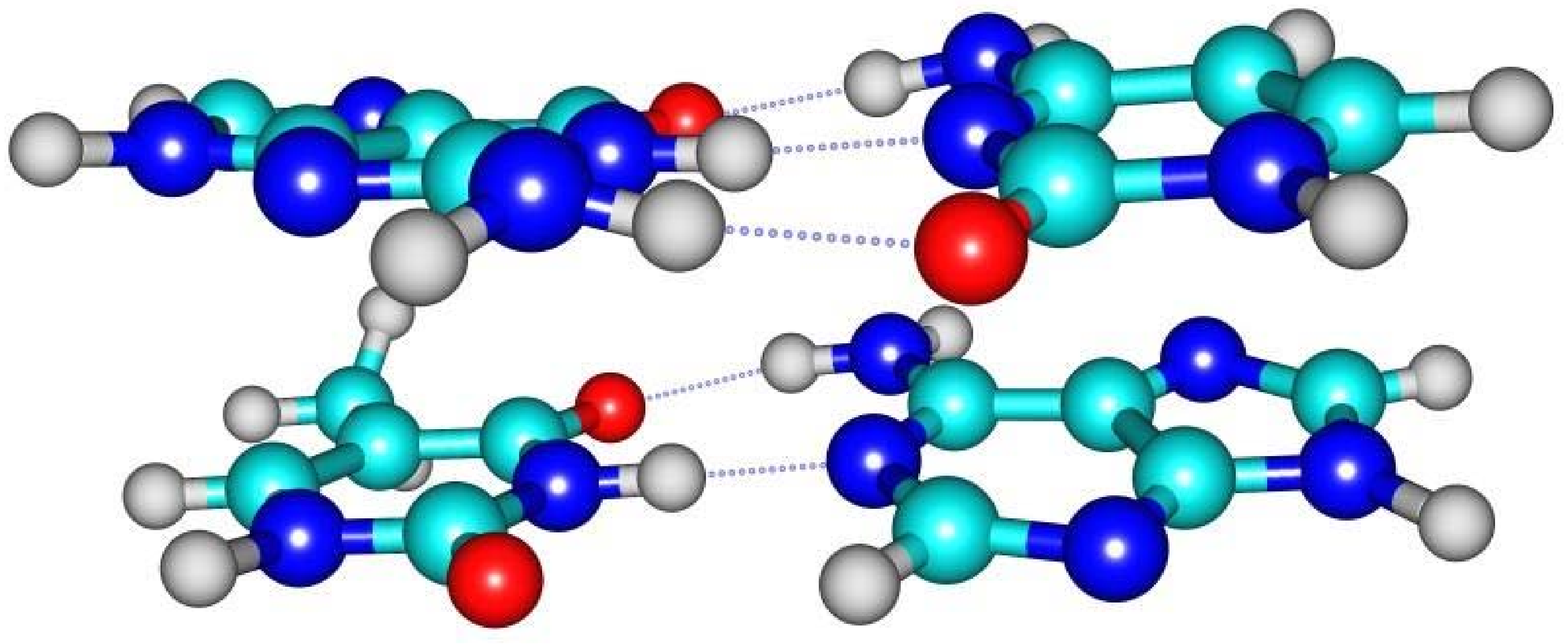}\\
(AC)(TG)\\ [2mm]
\includegraphics[width=5cm]{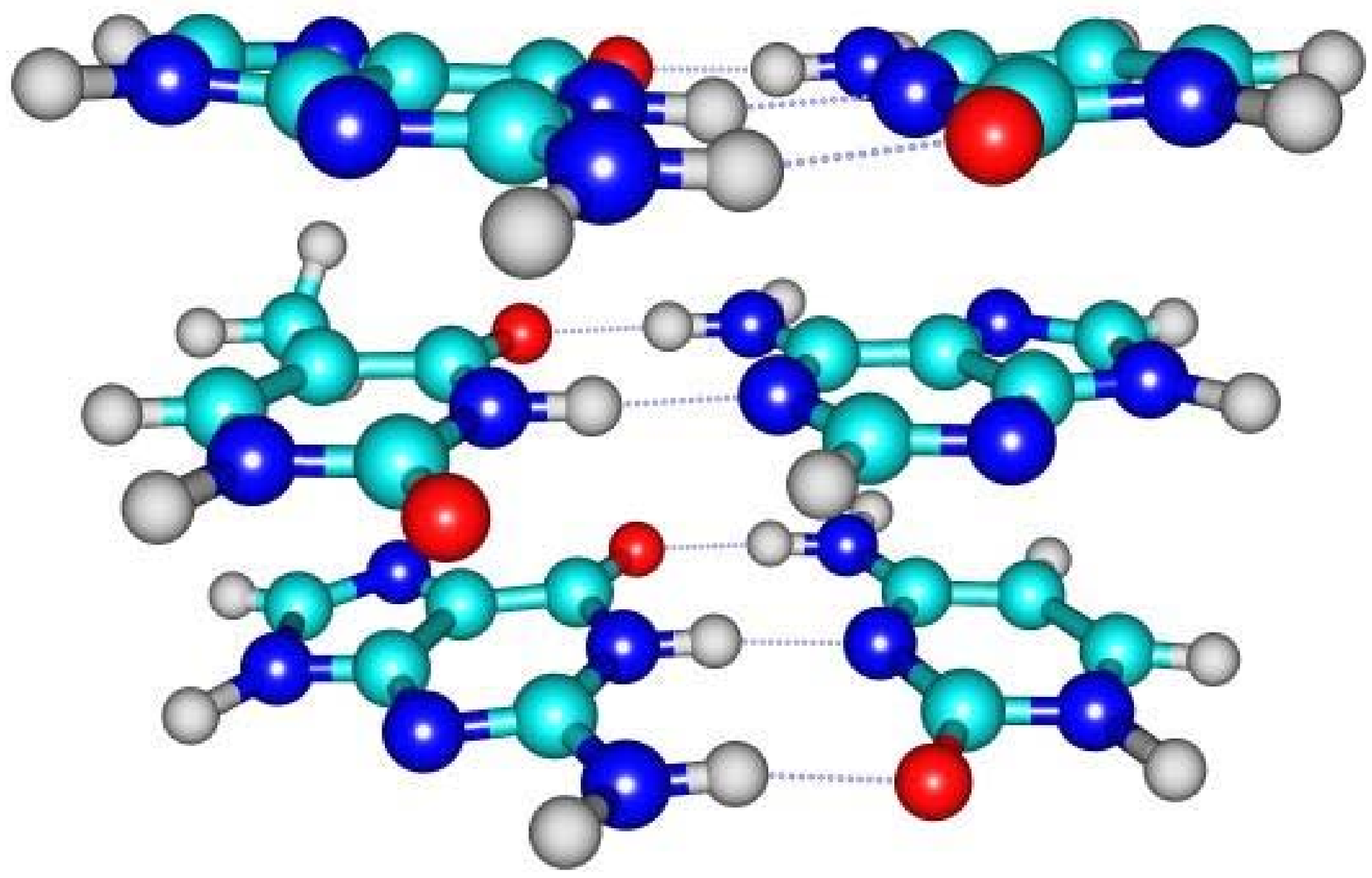}\\
(CAC)(GTG)\\ [2mm]
\includegraphics[width=5cm]{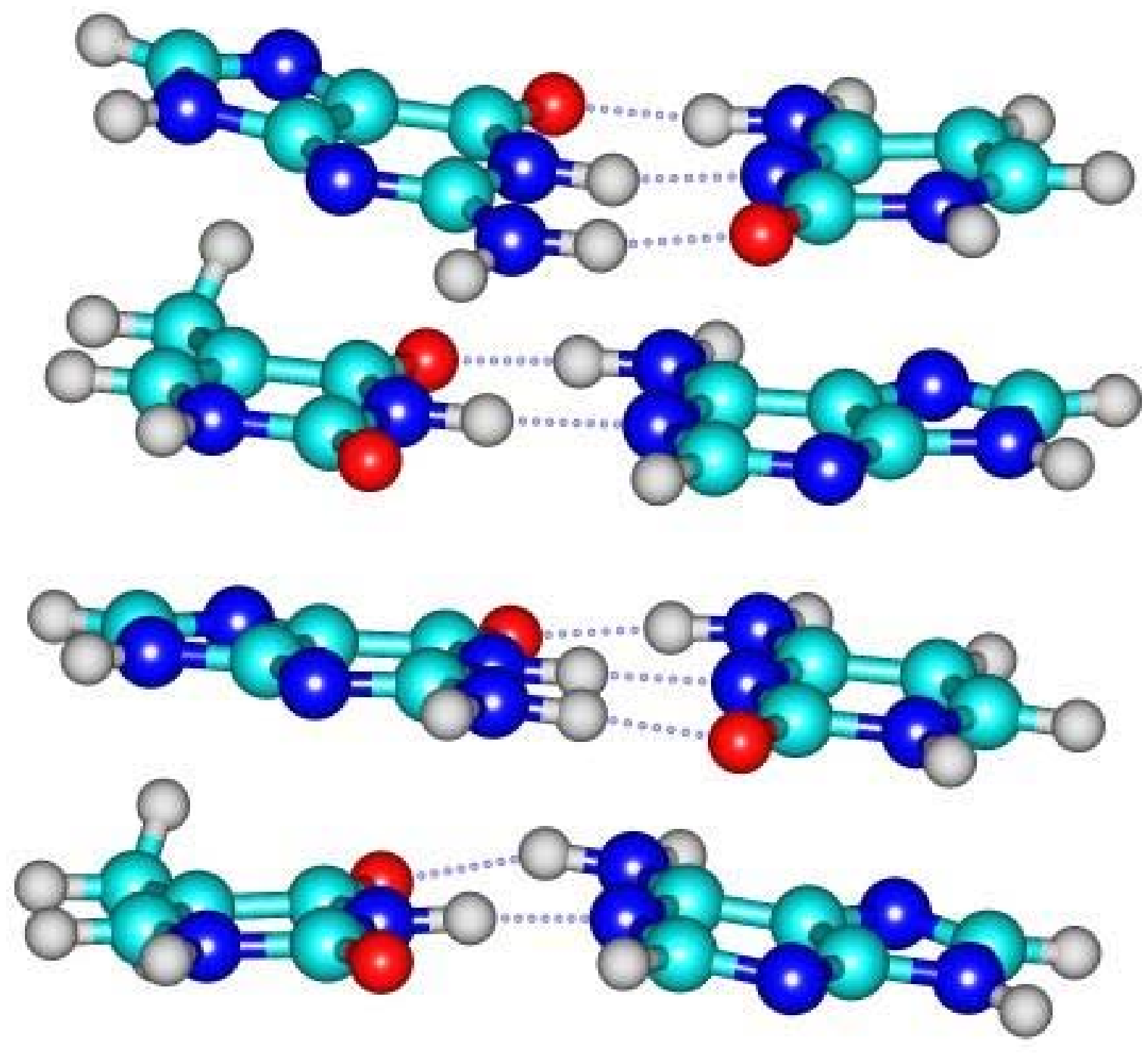}\\
(CACA(GTGT))\\
 \vskip-2mm\caption{ Optimized structures of the stacking
(AC) (TG) dimer, (CAC) (GTG) trimer, and (ACAC) (TGTG) tetramer of
nucleic base pairs }
\end{figure}

\begin{figure}
\includegraphics[width=\column]{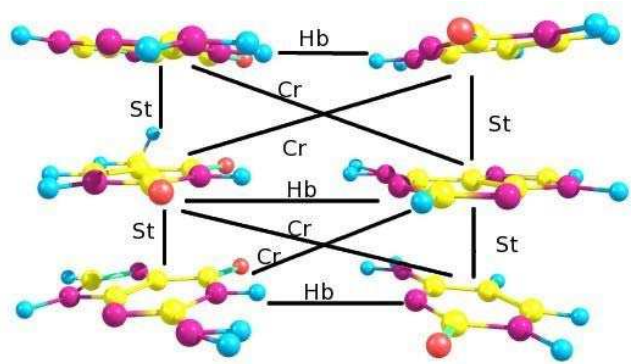}
\vskip-3mm\caption{ Calculated scheme of interactions in the trimer.
Assignations: H-b -- hydrogen bonds, St -stacking interactions, Cr
-- cross-interactions }
\end{figure}

\begin{table}[b]
\noindent\caption{Energies of horizontal interactions (kcal/mol) in
the stacking (AC)(TG) dimer, (CAC)(GTG) trimer, and (АCAC)(ТGTG)
tetramer calculated by various methods (6-31+G* basis set) and the
sums of the energies of H-bonds (kcal/mol) in isolated base
pairs}\vskip3mm\tabcolsep5.2pt
\noindent{\footnotesize\begin{tabular}{c c c c }
 \hline \multicolumn{1}{c}
{\rule{0pt}{9pt}Oligomer/} & \multicolumn{1}{|c}{MP2//MP2}&
\multicolumn{1}{|c}{MP2//M06-2X}&
\multicolumn{1}{|c}{Sums of H-bond}\\%
\multicolumn{1}{c}{Method}& \multicolumn{1}{|c}{}&
\multicolumn{1}{|c}{}&
\multicolumn{1}{|c}{energies}\\%
\hline%
Dimer & $-42.62$ & $-43.51$ & $-40.81$ \\%
Trimer &  & $-77.18$ &$-67.79$ \\%
Tetramer &  & $-94.72$ &$-81.62$ \\%
\hline
\end{tabular}}
\end{table}

To explain the contradiction revealed between the energies of horizontal interactions
and the sum of the energies of hydrogen bonds in base
pairs forming the oligomers, we used the idea of cross
interactions [18] and calculated the energies of cross interactions in the systems
under study. Cross interactions are induced by Coulomb, dispersion, and
inductive interactions and include interactions of every present
base with not only the nearest base, but with other ones in a stack
or in a helix (Figure 3).

So, the horizontal interaction energies must include H-bond energies and
cross interaction energies. The vertical interaction energies must
include stacking interaction and cross interaction energies.

\begin{table}[b]
\noindent\caption{Energies of interactions between components inside
the stacking (AC)(TG) dimer (MP2//MP2 method, 6-31+G* basis
set)}\vskip3mm\tabcolsep4.2pt
\noindent{\footnotesize\begin{tabular}{c c c c }
 \hline \multicolumn{1}{c}
{\rule{0pt}{9pt}No.} & \multicolumn{1}{|c}{Schemes$^a$ of
structures}& \multicolumn{1}{|c}{Interaction energies,}&
\multicolumn{1}{|c}{Sums$^g,$}\\%
\multicolumn{1}{c}{}& \multicolumn{1}{|c}{calculated}&
\multicolumn{1}{|c}{kcal/mol}&
\multicolumn{1}{|c}{kcal/mol}\\%
\hline%
1 & (A)$\leftrightarrow $(T)$^b$ & $-42.46^*$ & $-42.86$ \\%
 & (C)$\leftrightarrow $(G) &  & \\%
2 & (A)(T)$^c$ & $-12.80$ & \\%
 & (...)(...) &  & \\%
3 & (...)(...)$^c$ &$-26.31$ &\\%
 & (C)$\longleftrightarrow $(G) &  & \\%
4 &\underline{(A)(T)}$^d$ & $-12.02^{**}$ & $-12.39$\\%
 & (C)(G) &  &  \\%
5 & \underline{(A)}(...)$^e$ & $-3.88.$ & \\%
 & (C)(...) &  & \\%
6 & (...)\underline{(T)}$^e$ & $-4.77$ & \\%
 & (...)(G) & &\\%
7 & (A)(...)$^f$ & $-3.73.$ & \\%
 &(...)(G) &  & \\%
8 & (...)(T)$^f$ & $-0.01$ & \\%
 &(C)(...) &  & \\%
\hline
\end{tabular}
\raggedright{$^a$ Scheme of the (AC)(TG) stacking dimer in this
Table is\,$\begin{array}{c}
                                                            {\rm (A)(T)} \\
                                                            {\rm (C)(G)} \\
                                                          \end{array}\!\!\!\!.$\\
$^b$ Scheme $\begin{array}{c}
  ({\rm A})\leftrightarrow ({\rm T}) \\
  ({\rm C})\leftrightarrow ({\rm G}) \\
\end{array}$ means the breaking of the horizontal inte- raction in the (AC)(TG) stacking
dimer.\\
$^{c }$ Schemes $\begin{array}{c}
                 (A)\leftrightarrow (T) \\
                 (...)(...) \\
               \end{array}$
and $\begin{array}{c}
     (...)(...) \\
     ({\rm C})\leftrightarrow ({\rm G}) \\
   \end{array}$
show the breaking of the H-bond interaction in AT and CG pairs in
the (AC)(TG) stacking dimer. Geometries of these AT and~ GC~ pairs~
are~
identical~ to ones in the studied dimers.\\
$^d$ Scheme $\begin{array}{c}
               \underline{\rm( A)( T)}\\
               $(C)(G)$ \\
             \end{array}$
shows the breaking of the stacking interaction between AT and GC pairs
in (AC)(TG) stacking dimer.\\
$^e$ Schemes~ $\begin{array}{c}
                \underline{\rm(A)}(...) \\
                ({\rm C})(...) \\
              \end{array}$~
and~ $\begin{array}{c}
     (...)\underline{\rm(T)} \\
     (...)({\rm G}) \\
   \end{array}$
show the breaking of the sta- cking interactions in the stacking AC~
and~ TG dimers as a compo- nent of the (AC)(TG) stacking dimer of
base pairs.~ The~ geomet- ries~ of~ these~ AC~ and~ TG~ dimers~ are~
identical~ to~ ones~ in~
the (AC)(TG) dimer.\\
$^{f}$ Schemes $\begin{array}{c}
  ({\rm A})(...) \\
  (...)({\rm G}) \\
\end{array}$ and $\begin{array}{c}
                (...)({\rm T}) \\
                ({\rm C})(...) \\
              \end{array}$
show the breaking of cross in- teractions in the (AC)(TG) stacking
dimer
between A and G and between T and C, respectively.\\
$^{g}$ Sums of all possible interactions (H-bonds, stacking and
cross interactions) of components are shown. }

}
\end{table}

\subsection{Dimer}

Table 4 presents the calculation data on the energy of horizontal
interaction in the dimer of base pairs, energies of hydrogen bonds
in base pairs in the dimer configuration, energies of the vertical
interaction between AT and CG base pairs, energies of stacking
interactions in the dimers of bases, being a part of the base pair dimer,
as well as the energies of interactions between bases, being a part of
different pairs (cross energies). In all these calculations, the
geometry of bases was taken into account, which they possessed in
the base pair dimer composition. As stated above, the dimer geometry
was optimized at MP2 level of theory.

The following designations are accepted in Table 4. The second
column (lines {\it 1} and {\it 4}) presents the schemes of relative
positions of nucleic acid bases -- adenine (A), thymine (T), guanine
(G), and cytosine (C) in the calculated stacking dimer of nucleic
acid base pairs. This column also shows the schemes of relative
positions of nucleic bases in (A)(T), (G)(C) base pairs, in stacking
dimers of (A)(C) and (T)(G) pairs, and in cross pairs (A)(G) and
(T)(C). Horizontal arrows (line 1) mark the horizontal interaction
breakdown in the stacking dimer of base pairs. Horizontal arrows in
lines 2 and 3 point to the opening of hydrogen bonds in (A)(T) and
(G)(C) base pairs. Underlining in schemes (lines 4, 5, 6) shows the
stacking interaction failure in the dimers. The third column
presents the interaction energies calculated for the structures
shown in column 2. Column 4 gives the summation results for hydrogen
bond energies and cross interaction energies (line 1) and for
stacking and cross interaction energies (line 4) in the calculated
base pair dimer. It should be noted that the energy of hydrogen
bonds in AT and GC pairs being a part of the dimer is lower than
that of hydrogen bonds in isolated AT and GC pairs, which is related
to the different geometries of these pairs in the dimer composition
and in the free state.

As Table 4 shows (column 3), the energy of horizontal interaction in
the dimer is by 3.35 kcal/mol higher than the arithmetic sum of the
energies of hydrogen bonds in AT and CG base pairs in the dimer. The
energy of vertical interaction in the dimer between AT and CG pairs
is more than the arithmetic sum of the interaction energies in the
dimers of AC and TG bases, being a part of the stacking base pair
dimer.

As is noted above, this difference is probably caused by the cross
interactions. Really, the calculations of the interaction energies
between bases lying in different planes (cross energies) result in
the values of $-3.73$ kcal/mol and $-0.01$ kcal/mol for AG and TC
pairs, respectively. Taking these interactions into account allows
one to reach a rather good agreement between the horizontal and
vertical interaction energies and the sums of energies between
monomers in the stacking dimer (most right column in Table 4).
Differences between the total energy calculated for horizontal
interactions in the dimer and the arithmetic sum of hydrogen bonds
in base pairs and of cross interactions are within 1{\%}, similar to
the difference being about 3{\%} for the vertical interaction.

So, the presence of the cross interaction explains a rise of the
horizontal and vertical interaction energies in the stacking dimer,
in comparison with the sum of the energies of hydrogen bonds in base
pairs and the sum of the energies of stacking interactions in the
dimers of bases.

\begin{table*}[!]
\vspace*{3mm} \noindent\caption{Energies of interactions between
components inside the (CAC)(GTG) trimer (MP2//M06-2X method, 6-31+G*
basis set)}\vskip3mm\tabcolsep6.2pt
\noindent{\footnotesize\begin{tabular}{c c c c| cccc}
 \hline \multicolumn{1}{c}
{\rule{0pt}{9pt}No.} & \multicolumn{1}{|c}{Schemes of structures}&
\multicolumn{1}{|c}{Interaction energies,}& \multicolumn{1}{|c}{Sums
$^b,$}&\multicolumn{1}{|c} {\rule{0pt}{9pt}No.} &
\multicolumn{1}{|c}{Schemes of structures}&
\multicolumn{1}{|c}{Interaction energies,}&
\multicolumn{1}{|c}{Sums $^b,$}\\%
\multicolumn{1}{c}{}& \multicolumn{1}{|c}{calculated $^a$}&
\multicolumn{1}{|c}{kcal/mol}&
\multicolumn{1}{|c}{kcal/mol}&\multicolumn{1}{|c}{}&
\multicolumn{1}{|c}{calculated $^a$}& \multicolumn{1}{|c}{kcal/mol}&
\multicolumn{1}{|c}{kcal/mol}\\%
\hline%
1 & (C)$\longleftrightarrow $(G) & $-77.18$ & $-76.46^c$ &9 & (...)\underline{(G)} & $-4.17$ &\\%
 & (A)$\longleftrightarrow $(T) &  &$-77.16^d$ & & (...)(T) & &\\%
 & (C)$\longleftrightarrow $(G) &  & && (...)(...) &  &\\%
2 & (C)$\longleftrightarrow $(G) & $10^3$ &0.8170 &10 &(...)(...)& $-4.19$ &\\%
 & (...)(...) &  &&&  (...)\underline{(T)}& &   \\%
 & (...)(...) &  &&& (...)(G) &  & \\%
3 &(...)(...) & $-26.61$ & &11 & (...)(...) & $-4.47$ & \\%
 & (...)(...) &  &  && (A)(...) &  &\\%
 & (C)$\longleftrightarrow $(G) & 10.0 &0.87273 & & (...)(G) & &\\%
4 & (...)(...) & $-13.11$ & &12 & (...)(G) & $-4.39$ & \\%
 & (A)$\longleftrightarrow $(T) &  & & &(A)(...) &  &\\%
 & (...)(...) & && & (...)(...) &  & \\%
5 & \underline{(C)(G)} & $-11.00$ &$-11.70$ &13 & (...)(G)$^e$ & $-9.38$ &\\%
 &(A)(T) &  & && (A)(...) & &\\%
 & (C)(G) &  &  && (...)(G) &  &\\%
6 & (C)(G) & $-10.98$ &$-11.72$ &14 & (C)(...)$^f$ &$-1.44$ &\\%
 & \underline{(A)(T)} &  && & (...)(T) &  &  \\%
 & (C)(G) &  & & &(C)(...) &  &\\%
7 & \underline{(C)}(...) &$-2.51$ &&15 & (...)(...) & $-0.63$ & \\%
 & (A)(...) &  && & (...)(T) &  &  \\%
 &(...)(...) &  &&& (C)(...) &  & \\%
8 & (...)(...) & $-2.43$ & &16 & (C)(...) & $-0.63$ &  \\%
 & \underline{(A)}(...) &  && & (...)(T) & & \\%
 & (C)(...)&  & && (...)(...) &  & \\%
\hline
\end{tabular}
\raggedright{ $^{a }$Scheme of the (CAC)(GTG) stacking trimer in
this table is \!$\begin{array}{c}
                                                               ({\rm C})\leftrightarrow ({\rm G}) \\
                                                               ({\rm A})\leftrightarrow ({\rm T}) \\
                                                               ({\rm C})\leftrightarrow ({\rm G}) \\
                                                             \end{array}$\!\!.
Other~ designations~ in~ this~ Table~ are~ similar~ to~ those~ in
Table\,4. Horizontal interaction in this Table is between (CAC) and
(GTG) stacking trimers of nucleic bases. Stacking interactions are
between the upper CG pair and (AC)(GT) components and between
(CA)(GT) and the down CG pair.

$^{b}$Sums of all possible interactions (H-bonds, stacking and cross
interactions) of the components.

$^{c}$Sum of H-bond interactions with using separate GA, AG, CT and
TC (lines 11, 12, 14, 15) cross interactions.

$^{d}$Sum of H-bond interactions together with GAG (line13) and CTC
(line 16) cross interactions.

} }\vspace*{-1mm}
\end{table*}

\subsection{Interactions in the trimer and the tetramer}

The same situation is observed both for the trimer and the tetramer.
Table 5 presents the energies of horizontal interaction in the
trimer of base pairs, energies of hydrogen bonds in base pairs in
the trimer, energies of vertical interactions between AT and CG base
pairs, and energies of stacking interaction in the dimers of bases,
being a part of the trimer, as well as the energies of interactions
between bases forming different pairs (cross interaction energies).
In all these calculations, the geometry of bases was used, which
they have in the trimer structure. In addition, Table 5 presents the
results of summation of hydrogen bond energies (rightmost column)
and of the energies of AG and TC cross interactions in the trimer
(above) and the results of summation of stacking interaction
energies in the dimers of AC and TG bases forming the studied
stacking trimer of (CAC)(GTG) base pairs and the energies of AG and
TC cross interactions.

It should be noted that the energies of hydrogen bonds in AT and GC
pairs in the trimer differs both from those of hydrogen bonds in the
dimer and in isolated AT and GC pairs, which is determined by a
difference in the geometries of these pairs in the trimer, dimer,
and in the free state (Table 2). As is seen from Table 5, the
calculations revealed that the energy of horizontal interaction in
the stacking trimer exceeds the arithmetic sum of the hydrogen bond
energies in isolated AT and GC pairs. The excess makes up $-10.84$
kcal/mol. Energies of the vertical interactions in the trimer
between AT and CG pairs are larger than the arithmetic sums of
interaction energies in the dimers of bases -- AC and TG being a
part of the stacking trimer composition.

The calculations of interaction energies of the bases placed in
various planes (cross interaction energies) resulted in $-4.39$
kcal/mol and $-4.47$ kcal/mol for GA and AG pairs and $-0.63$
kcal/mol for CT and TC pairs, respectively. There is some difference
of interaction energies for reciprocal GA and AG pairs. Table 5 also
presents the energies of cross interactions between bases (adenine
and guanine) in the GAG triad and cytosine and thymine in the CTC
triad of the stacking trimer under study. Just taking these cross
interactions into account allows us to reach a very good agreement
between the total energy of horizontal interactions and the sum of
the energies of interactions between monomers (nucleic acid bases)
in the trimer.

Table 5 shows the energies of vertical interactions between the
upper GC and middle AT pairs and between middle AT and down GC
pairs, which are only by about 0.7 kcal/mol lower than the summation
results for stacking and cross interactions of components in these
structures. Thus, as in the dimer case, the consideration of the
cross interaction explains a rise of the energy of horizontal and
vertical interactions in the base pair trimer in comparison with the
sum of the energies of hydrogen bonds in base pairs and with the sum
of stacking interactions in dimers of bases, being the part of the
trimer.\looseness=1

Let us consider the interaction energies for the tetramer. Table 1
from Supplement presents the calculation results (column 3) for the
total energy of horizontal interaction in the tetramer,  energies of
hydrogen bonds in base pairs being a part of the tetramer, and
energies of paired interactions (stacking and cross interactions) in
the tetramer. As in the cases of dimer and trimer, the energies of
horizontal and vertical interactions exceed the arithmetic sum of
the energies of hydrogen bonds and stacking interactions in pairs
and the dimers of bases being in the tetramer. Nevertheless, the
consideration of the cross interactions in cross pairs and triads of
bases present in the tetramer allows us to ``compensate'' this
difference.

It is worth noting the variability of the energies of cross interactions in AG and TC pairs in
the tetramer. The energies of cross interactions
change from $-4.12$ kcal/mol to $-4.70$ kcal/mol and from $-0.42$
kcal/mol to $-0.81$ kcal/mol for AG and CT pairs, respectively. This
variability is related to the difference in the spatial structures
of complementary AT and GC base pairs (in particular, with the
difference in dihedral and propeller angles, which include AG and CT
``cross pairs'' (see Table~2)).

\begin{figure*}
\includegraphics[width=7.5cm]{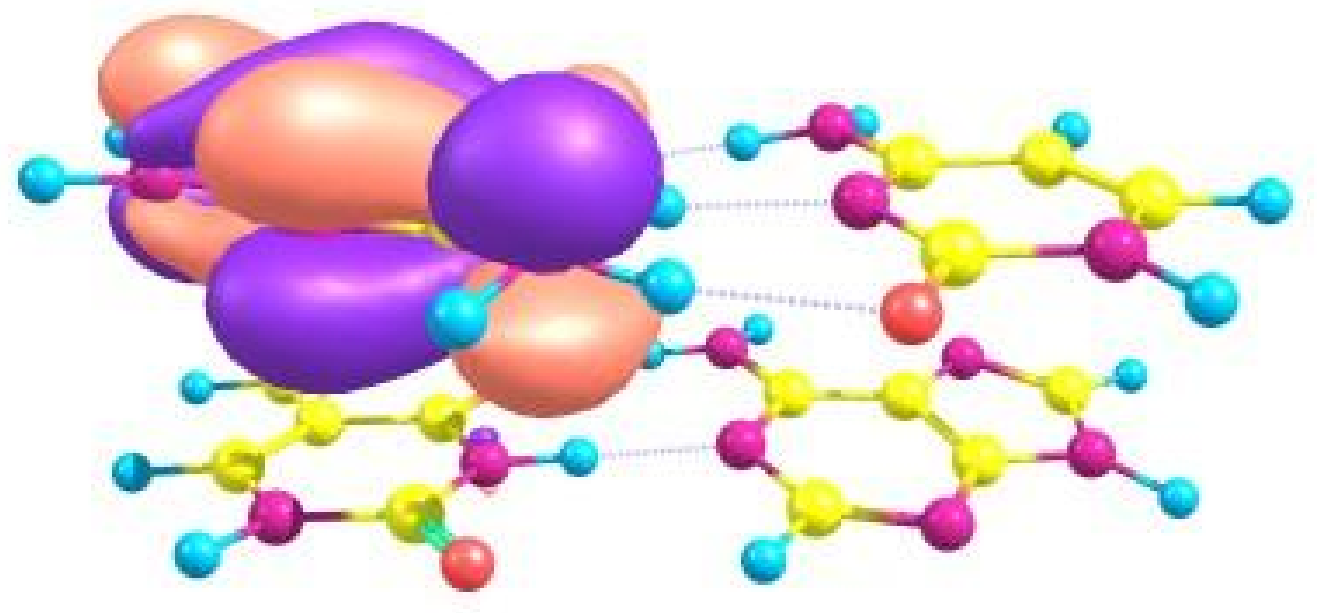}\hspace{1.5cm}\includegraphics[width=7.5cm]{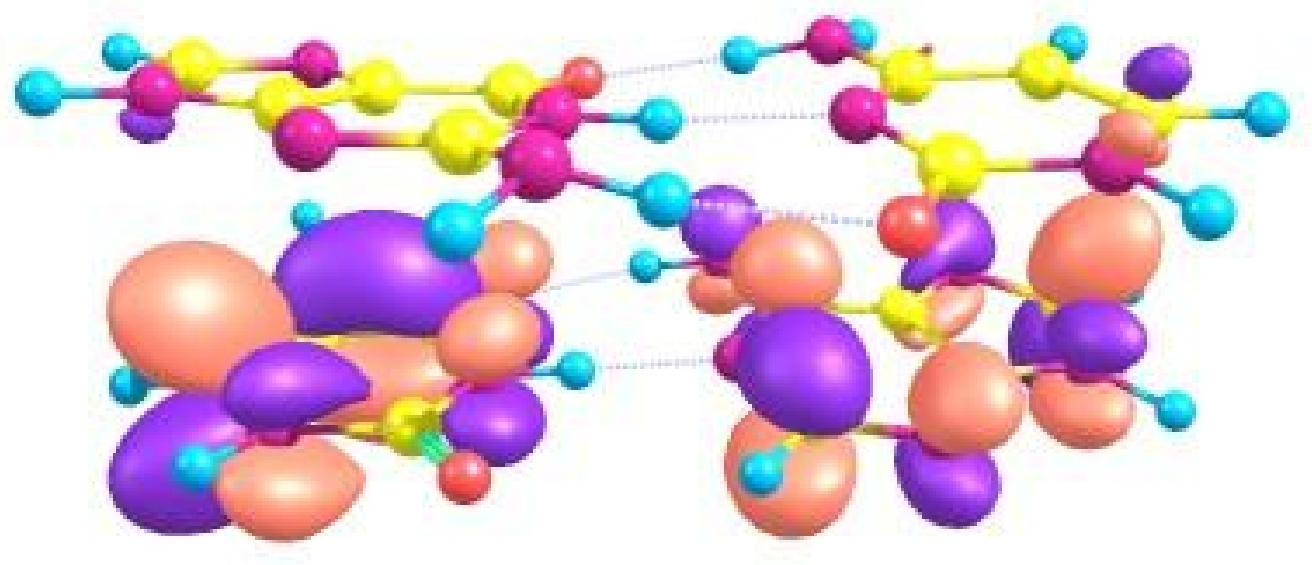}\\
{HOMO\hspace{8.5cm}LUMO+6} \vskip-3mm\caption{ Molecular orbitals
(HOMO and LUMO+6), between which the first electronic transition
occurs in the dimer }\vskip5mm
\end{figure*}

In addition, we calculated the energies necessary for the
``opening'' of end AT and GC base pairs, being a part of the
tetramer. The values of energies were $-23.36$ kcal/mol and $-37.49$
kcal/mol for AT and GC pairs, respectively. The values of energies
calculated directly for end pair openings are close to those of the
sum of the energies of hydrogen bonds in base pairs, as well as the
energies of cross and stacking interactions between the nearest
bases of nucleic pairs in the tetramer. The calculations of the
opening energy of the outer GC pair in the tetramer and the trimer
gave similar results: $-37.49$ kcal/mol and $-37.98$ kcal/mol for
the tetramer and the trimer, respectively.

The analysis of Tables 3--5 and Table 1 in Supplement allowed us to
reveal increasing the relative contribution of cross interactions
to the total horizontal interactions in the structures under study. This
contribution increases from 7.9{\%} for the dimer to 18.4{\%} for
the tetramer.

\subsection{Charges}

The analysis of the electronic structure of oligomers (in
particular, the analysis of the electronic density distribution) is
of special interest. The calculations of the charges on atoms (Table
6) showed that the sum of charges in stacks of (AC)and(TG), (CAC)
and (GTG), (ACAC) and (TGTG) bases forming the oligomers under study
is not equal to 0 and comes to 0.115, 0.229, and 0.243 for the
dimer, trimer, and tetramer, respectively. It is a manifestation of
the charge transfer in the horizontal direction in the short
stacking oligomers.

\begin{table}[t]
\noindent\caption{Total charges of base stacks in the (AC)(TG) dimer,
(CAC)(GTG) trimer, and (ACAC) (TGTG)
tetramer}\vskip3mm\tabcolsep19.2pt
\noindent{\footnotesize\begin{tabular}{c c }
 \hline \multicolumn{1}{c}
{\rule{0pt}{9pt}Stacks of nucleobases} & \multicolumn{1}{|c}{Total charges in stacks of}\\%
\multicolumn{1}{c}{in oligomers}& \multicolumn{1}{|c}{nucleobases}\\%
\hline%
(AC) & 0.115 \\%
(TG) &$-0.115$ \\%
(CAC) &0.229 \\%
(GTG) &$-0.229$ \\%
(ACAC) &0.243\\%
(TGTG) &$-0.243$ \\%
\hline
\end{tabular}}
\end{table}

\subsection{Excited states}

The analysis of excited states in the structures is of significant
interest as well. The calculations of the excited state energies in
the stacking dimer and trimer revealed a possibility for a
low-intensity band to exist in the long-wave region of the spectrum
(Table 7). This corresponds to our previous suppositions [19] on
possible charge-resonance interactions in crystals and dimers of
nucleic acid bases, as well as to the data of work [20]. The
analysis of molecular orbitals in the dimer showed (Fig. 4) that
this low-frequency transition has to take place between the
molecular orbital localized on guanine and the molecular orbital
localized on the neighboring pair of bases (thymine and adenine).
Thus, we deal with the electronic transition between two systems of
molecules being placed one over another. Moreover, in the trimer
case, we can observe several low-intensity electronic transitions in
the low-energy region of the spectrum.

\begin{table}[t]
\noindent\caption{Parameters of dimer (left) and trimer (right)
excited states calculated by the TD DFT method on MP2 level of
theory calculated geometries}\vskip3mm\tabcolsep1.9pt
\noindent{\footnotesize\begin{tabular}{c c c c c c c }
 \hline \multicolumn{1}{c}
{\rule{0pt}{9pt}Excited } & \multicolumn{1}{|c}{Excited }&
\multicolumn{1}{|c}{Wave}& \multicolumn{1}{|c}{Oscil. }&
\multicolumn{1}{|c}{Excited}& \multicolumn{1}{|c}{Wave }&
\multicolumn{1}{|c}{Oscil.}\\%
\multicolumn{1}{c} {state } & \multicolumn{1}{|c}{ state }&
\multicolumn{1}{|c}{length,}& \multicolumn{1}{|c}{strength }&
\multicolumn{1}{|c}{ state}& \multicolumn{1}{|c}{length, }&
\multicolumn{1}{|c}{strength}\\%
\multicolumn{1}{c} { } & \multicolumn{1}{|c}{energies,}&
\multicolumn{1}{|c}{nm}& \multicolumn{1}{|c}{ }&
\multicolumn{1}{|c}{energies,}& \multicolumn{1}{|c}{nm }&
\multicolumn{1}{|c}{}\\%
\multicolumn{1}{c}{}& \multicolumn{1}{|c}{eV, dimer}&
\multicolumn{1}{|c}{}& \multicolumn{1}{|c}{}&
\multicolumn{1}{|c}{eV,  trimer}&
\multicolumn{1}{|c}{}& \multicolumn{1}{|c}{}\\%
\hline%
1 & 4.70 & 263.4 & 0.0040 & 4.60 & 269.1 & 0.0030 \\%
2 & 4.86 & 254.9 & 0.012 & 4.64 & 267.1 & 0.0034 \\%
3 & 4.91 & 252.6 & 0.038 & 4.69 & 264.2 & 0.0033 \\%
4 & 4.94 & 250.1 & 0.016 & 4.75 & 260.7. & 0.0479 \\%
5 & 5.05 & 245.3 & 0.001 & 4.76 & 260.3 & 0.0004 \\%
6 & 5.12 & 242.3 & 0.008 & 4.81 & 258. & 0.0401 \\%
7 & 5.14 & 241.1 & 0.050 & 4.82 & 257.3 & 0.0051 \\%
8 & 5.17 & 239.7 & 0.282 & 4.82 & 257.0 & 0.0132 \\%
9 & 5.24 & 236.3 & 0.101 & 4.95 & 250.0 & 0.001 \\%
10 & 5.36 & 231.2 & 0.243 & 4.99 & 248.21 & 0.0558 \\%
\hline
\end{tabular}}
\end{table}

\begin{table}[t]
\vskip2mm \noindent\caption{Energies of interactions between
components inside a tetramer (MР2//M06-2X computation
method)}\vskip3mm\tabcolsep4.2pt
\noindent{\footnotesize\begin{tabular}{c c c |ccc}
 \hline \multicolumn{1}{c}
{\rule{0pt}{9pt}No.} & \multicolumn{1}{|c}{Schemes}&
\multicolumn{1}{|c}{Interaction}&\multicolumn{1}{|c}
{\rule{0pt}{9pt}No.} & \multicolumn{1}{|c}{Schemes }&
\multicolumn{1}{|c}{Interaction}\\%
\multicolumn{1}{c} {} & \multicolumn{1}{|c}{structures}&
\multicolumn{1}{|c}{energies,}&\multicolumn{1}{|c} {} &
\multicolumn{1}{|c}{structures}&
\multicolumn{1}{|c}{energies,}\\%
\multicolumn{1}{c}{}& \multicolumn{1}{|c}{calculated}&
\multicolumn{1}{|c}{kcal/mol}&\multicolumn{1}{|c}{}&
\multicolumn{1}{|c}{calculated}& \multicolumn{1}{|c}{kcal/mol}\\%
\hline%
1 & (A)$\longleftrightarrow $ (T) & $-94.72$&9 & (...)(T) & 0.42    \\%
 & (C)$\longleftrightarrow $ (G) & & & (C)(...) &   \\%
 & (A)$\longleftrightarrow $ (T)&  & & (...)(...) & \\%
 & (C)$\longleftrightarrow $ (G) & && (...)(...) &   \\%
2 & (...)(...) &$-26.79$ &10 &(...)(T) & $-0.81$ \\%
 & (C)$\longleftrightarrow $ (G) &   & & (C)(...) &\\%
 &(...)(...) &  & & (...)(T) &  \\%
 & (...)(...) & & &(...)(...) &\\%
3 & (...)(...) & 26.44 &11 & (...)(...) & $-0.76$  \\%
 &(...)(...) & & & (C)(...) &   \\%
 & (...)(...) & & & (...)(T) &   \\%
 & (C)$\longleftrightarrow $ (G) &  && (C)(...) &\\%
4 & (A)$\longleftrightarrow $ (T) & $-12.03$&12 & \underline{(A)}(...) &$-2.59$  \\%
 & (...)(...) & && (C)(...) &  \\%
 & (...)(...) & &&(...)(...) &\\%
 & (...)(...) &  & & (...)(...) &  \\%
5 &(...)(...) & $-12.04$ &13 & (...)(...) & $-2.446$ \\%
 & (...)(...) & &&(...)(...) & \\%
 & (A)$\longleftrightarrow $ (T) & & & \underline{(A)}(...) &   \\%
 &(...)(...) & && (C)(...) &  \\%
6 & (A)(...) & $-4.70$ &14 & (...)(...) & $-3.64$ \\%
 & (...)(G) & & & (...)(...) &   \\%
 & (...)(...) &  && (...)\underline{(T)} &\\%
 & (...)(...) &  && (...)(G) &   \\%
7 & (...)(...) &$-4.35$ &15 &(...)\underline{(T)} & $-2.55$\\%
 & (...)(...) &&& (...)(G) &  \\%
 &(A)(...)  & && (...)(...) &  \\%
 & (...)(G) & & &(...)(...) &  \\%
8 & (...)(...) & $-4.12$  \\%
 &(...)(G) &   \\%
 & (A)(...) &   \\%
 & (...)(...) &   \\%
\hline
\end{tabular}
\vskip1mm\raggedright{Assignations are the same as in Tables 4 and
5.

}}\vspace*{-4mm}
\end{table}

\section{Conclusions}

1. The {\it ab initio} calculations of nanoparticles of nucleic
acids components -- a dimer and a trimer of nucleotides and a
stacking trimer and a stacking tetramer of nucleic bases have been
carried out for the first time.

2. The calculations performed revealed peculiarities of the structure and
the interaction energy of nanoparticles:

\noindent -- Cross interactions play a considerable role in
the stabilization of the structures. The contribution of cross
interactions into horizontal interactions grows with the
lengthening of the oligomer.

\noindent -- Internal pairs of bases in a trimer and a tetramer of base
pairs are more planar in comparison with external ones.

\noindent -- Nanoparticle components get electric charges in
nanoparticles.

\noindent -- Low-intensity long-wave bands can appear in the electronic
spectra of nanoparticles.

\vskip3mm The authors thank Dr. S. Stepanian and Dr. R. Zybatyuk for
their help in calculations. We appreciate the Institute of
Cybernetics of NASU, Institute of Theoretical Physics of NASU,
Institute of Single Crystals of NASU, Moscow State University, and
Institute for Low Temperature Physics and Engineering of NASU for
softwares and the computer time given.

\section{Supplement}
See Table 8.

\vspace*{-3mm}
\rezume{%
МОЛЕКУЛЯРНА СТРУКТУРА ТА ВЗАЄМОДІЇ\\ У~~~ НАНОЧАСТИНКАХ~~~ КОМПОНЕНТІВ\\
НУКЛЕЇНОВИХ КИСЛОТ: НЕЕМПІРИЧНІ\\ РОЗРАХУНКИ}{Ю.В. Рубін, Л.Г.
Белоус} {Самоасоціати компонентів нуклеїнових кислот (стекинг
тримери та тетраметри основ нуклеїнових кислот) та короткі фрагменти
нуклеїнових кислот є наночастинками (їхні лінійні розміри більше
10~{\AA}). Сучасні квантово-механічні методи та програми дозволяють
здійснювати розрахунки систем, що містять 150--200 атомів, із
достатньо великим базисом (наприклад, 6-31G*). Метою роботи є
виявлення особливостей молекулярної та електронної структур,
енергетичних характеристик наночастинок компонентів нуклеїнових
кислот.

Нами виконано неемпіричні розрахунки молекулярної структури та
взаємодій в стекинг димері та тримері пар нуклеотидів, а також в
стекинг димері, тримері та тетраметрі пар основ нуклеїнових кислот.

Проведені розрахунки молекулярної структури димеру та тримеру пар
нуклеотидів показали, що міжплощинна відстань в структурах, що
вивчалися, в середньому дорівнювала 3,2~{\AA}, кут спіральності в
тримері -- $30^{\circ} $, відстань між атомами фосфору в сусідніх
ланцюгах 13,1~{\AA}. Також розраховані енергії горизонтальних
взаємодій у цих димерах та тримерах.

Ми зробили детальний аналіз міжплощинних відстаней і кутів між
основами та їх парами у стекинг димері, тримері та тетраметрі пар
основ нуклеїнових кислот. Аналіз горизонтальних і вертикальних
взаємодій в розрахованих коротких олігомерах пар основ нуклеїнових
кислот показав значну роль перехресних взаємодій у стабілізації
вивчених структур. Внесок перехресних взаємодій в горизонтальні
взаємодії збільшується з подовженням олігомеру. Показано, що у
складі наночастинок їхні компоненти отримують електричний заряд. В
електронних спектрах наночастинок можуть з'явитись довгохвильові
малоінтенсивні смуги.}

\end{document}